\documentclass[showpacs,twocolumn,aps,prb]{revtex4-1}

\usepackage{amsmath}
\usepackage{amssymb}
\usepackage{graphicx}
\usepackage{dcolumn}
\usepackage{bm}

\usepackage{bm}
\usepackage{graphicx}
\usepackage{ulem}

\def\be{\begin{equation}}
\def\ee{\end{equation}}
\def\ber{\begin{eqnarray}}
\def\eer{\end{eqnarray}}
\def\bwt{\begin{widetext}}
\def\ewt{\end{widetext}}

\def\e{{\varepsilon}}
\def\bt{\textbf}

\def\e{\varepsilon}
\def\o{\omega}

\begin{document}

\title{Plasmonic excitations in Coulomb coupled $N$-layer graphene structures}
\author{J.-J. Zhu}
\affiliation{Department of Physics, University of Antwerp, Groenenborgerlaan 171, B-2020 Antwerpen, Belgium}
\author{S. M. Badalyan}
\email{Samvel.Badalyan@ua.ac.be}
\affiliation{Department of Physics, University of Antwerp, Groenenborgerlaan 171, B-2020 Antwerpen, Belgium}
\author{F. M. Peeters}
\affiliation{Department of Physics, University of Antwerp, Groenenborgerlaan 171, B-2020 Antwerpen, Belgium}

\begin{abstract}
We study Dirac plasmons and their damping in spatially separated $N$-layer graphene structures at finite doping and temperatures. The plasmon spectrum consists of one optical excitation with a square-root dispersion and $N-1$ acoustical excitations with linear dispersions, which are undamped at zero temperature within a triangular energy region outside the electron-hole continuum. For any finite number of graphene layers we have found that the energy and weight of the optical plasmon increase in the long wavelength limit, respectively, as square-root and linear functions of $N$. This is in agreement with recent experimental findings. 
With an increase of the number of multilayer acoustical plasmon modes, the energy and weight of the upper lying branches also exhibit an enhancement with $N$. This increase is strongest for the uppermost acoustical mode so that its energy can exceed at some value of momentum the plasmon energy in an individual graphene sheet. Meanwhile, the energy of the low lying acoustical branches decreases weakly with $N$ as compared with the single acoustical mode in double-layer graphene structures. Our numerical calculations provide a detailed understanding of the overall behavior of the wave vector dependence of the optical and acoustical multilayer plasmon modes and show how their dispersion and damping are modified as a function of temperature, interlayer spacing, and inlayer carrier density in (un)balanced graphene multilayer structures.
\end{abstract}


\maketitle

\section{Introduction}

Graphene\cite{graphene1,graphene2}, a single-atom sheet of graphite, has attracted considerable attention due to its zero band gap electronic structure with massless linear and chiral dispersion of charge carriers \cite{Geim2009,Novo2011}. These unique characteristics lead to numerous of extraordinary new phenomena in graphene such as the Klein tunneling \cite{Geim2006}, the ambipolar field effect \cite{graphene1,Chen2010}, the finite minimum conductivity  \cite{graphene2,Tan2007}, the ultra-high mobility \cite{Bolotin2008,Du2008}, and the anomalous quantum Hall effect \cite{graphene2,Kim2005}. With these novel properties \cite{Sarma2011,Cooper2012}, graphene presents itself as an extremely promising candidate-material for future generation of electronics \cite{geimnovo2007,geimmcd2007}. 

Graphene structures open a new arena also for studying fundamental many-body interaction phenomena, which can play also an important role in graphene based electronic devices \cite{Kotov2012}. Particularly, electron-phonon interaction phenomena have become a subject of active study in single-layer graphene structures in zero\cite{Park2007,Calandra2007,Tse2007,SMB2012ph} and finite magnetic fields \cite{Ando2007,Goerbig,Faugeras,SMB2012mph}. Substantial efforts have been directed towards the investigation of the linear response of doped graphene \cite{Principi2009} and of the charge density excitations \cite{Wunsch2006,Hwang2007,Barlas2007,DasSarma2009,Abedinpour2011} and of such complex quasiparticles as plasmarons \cite{Polini2008,Bostwick2010} and plasmon-phonon complexes \cite{Jablan2011}. Recently, the experimental realization of graphene double-layer structures coupled only via the Coulomb interaction\cite{Haug2008,Tutuc2011,Ponamarenko2011,Tutuc2012,Britnell2012,Gorbachev2012} has attracted substantial theoretical interest in studying the double-layer plasmon effects \cite{Hwang2009,Stauber2012,Profumo2012,SMB2012} and the frictional drag \cite{Rojo1999} in two spatially separated graphene layers
\cite{Tse2007D,Narozhny2007,Castro2011,Katsnelson2011,Hwang2011,Narozhny2012,Carega2012,SMB2012D,Matos2012} as powerful tools for probing interaction effects of massless Dirac fermions.

One of the key quantities for investigating plasmon properties in Fermi liquid theory \cite{GV} is the dynamical polarizability, which, together with the Coulomb interaction potential, determines the dielectric function of an electronic system and describes its dispersive and dissipative properties. The recent theoretical calculations of the bubble diagram of the Lindhard polarization function, $\Pi_{0}({\omega },q)$, in monolayer graphene \cite{Wunsch2006,Hwang2007,Barlas2007} show that for an arbitrary bosonic frequency, ${\omega }$, and momentum, $q$, its dependence on the carrier density, $n$, is weaker, $\Pi_{0}({\omega },q)\propto
\sqrt{n}$, than the dependence $\Pi_{0}({\omega },q)\propto n$ in conventional two dimensional electron systems with a parabolic energy dispersion. As a direct consequence, the frequency of charge density waves, ${\omega }_{p}$, in a quantum system of chiral massless Dirac fermions shows a unique carrier density scaling, ${\omega }_{p}\propto n^{1/4}$, which is in contrast with conventional two-dimensional plasma ${\omega }_{p}\propto n^{1/2}$.
The weak dependence of the plasmon frequency and its weight on the carrier doping level can be a limiting factor in the realization of effective control and tunability of graphene-based applications, which due to many novel properties hold a promise of completely new functionalities in optoelectronics and terahertz metamaterials \cite{Grigorenko2012,Fei2012,Fei2011,Ju2011}.

Recently, Ref.~\onlinecite{Yan2012nnano} demonstrates experimentally a new tunable photonic device based on graphene/insulator stacks, formed by depositing $N=5$ alternating wafer-scale graphene sheets and thin insulating layers.  Measurements \cite{Yan2012nnano} of light-plasmon coupling in such graphene multilayer structures (GMLS) show that distributing carriers into several graphene layers enhances effectively the plasmon frequency and the magnitude of extinction of the light transmission in comparison with that in single-layer graphene of the same total density. The measured frequency and weight enhancement of the in-phase optical plasmon mode follows, respectively, the square-root and the linear dependence on the number of graphene sheets in GMLS, {\it i.e.} $\sqrt{N}$ and $N$, which differs from the weak inlayer density dependence, $n^{1/4}$ and $n^{1/2}$, obtained respectively for the plasmon frequency and its weight both in single-\cite{Wunsch2006,Hwang2007,Barlas2007,DasSarma2009,Abedinpour2011} and double-layer \cite{Hwang2009,Stauber2012,Profumo2012,SMB2012} graphene structures. This effect is strikingly different from that in conventional two-dimensional systems in semiconductors where, if exchange and correlations effects \cite{SMB2007} are neglected, the in-phase plasmon mode in a symmetrically balanced double-layer system behaves similarly to that in an individual layer with the doubled density. The observed unusual behavior of the in-phase optical plasmon mode in GMLS has been ascribed \cite{Yan2012nnano} to the nonclassical plasmon frequency of massless Dirac fermions, which depends explicitly on the Planck constant $\hbar $.

\begin{figure}[t]
\includegraphics[width=\columnwidth]{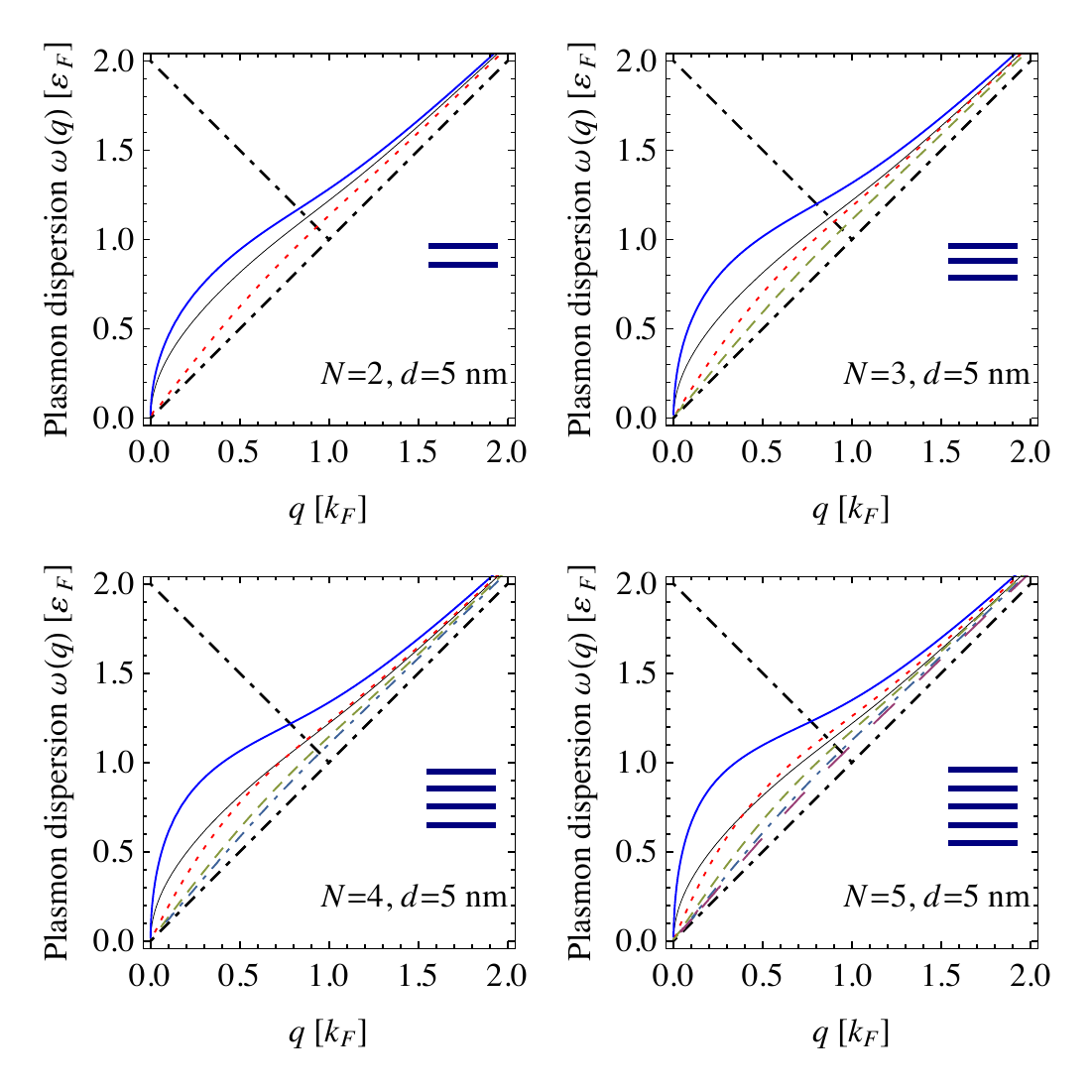}
\caption{(Color online) The plasmon dispersions at finite temperature $T=0.1T_{F}$ in GMLS consisting of $N=2,3,4$ and $5$ graphene monolayers (shown in bold parallel lines in each panel) for the interlayer separation $d=5 $ nm. The upper solid curve represents the optical plasmon mode with the square-root dispersion while lower dotted, dashed, dotted-dashed, and long-dashed curves represents acoustical branches of the plasmon spectrum with linear dispersions. The thin black line represents the plasmon mode in an individual graphene layer. The doping level in each layer corresponds to the inlayer carrier density $n_{i}=n_{0}=10^{12}$ cm$^{-2}$. The thick dot-dashed lines show the boundaries of inter- and intra-subband electron-hole continua.  
All quantities in this figure are dimensionless.}
\label{fig1}
\end{figure}

\begin{figure}[t]
\includegraphics[width=\columnwidth]{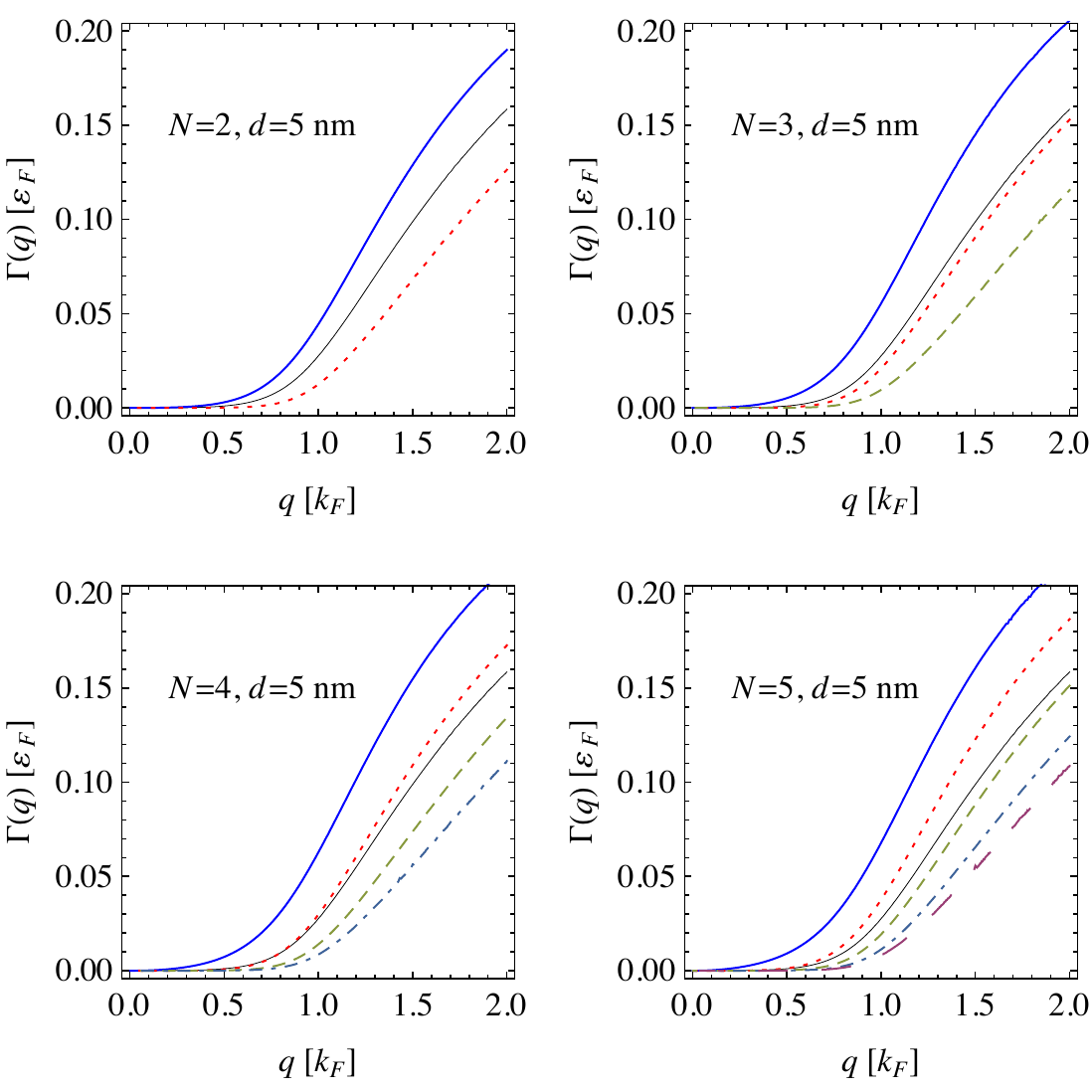}
\caption{(Color online) The damping function of multilayer plasmon modes at finite temperature $T=0.1T_{F}$ in GMLS consisting of $N=2,3,4$ and $5$ graphene monolayers. The curves plotted here corresponds to the plasmon branches represented in Fig.~\ref{fig1} with the same formatting. All system parameters are taken the same as in Fig.~\ref{fig1}.}
\label{fig2}
\end{figure}

In this paper we present a detailed study of the spectrum of multilayer optical and acoustical plasmon modes in GMLS, consisting of a {\it finite} number, $N$, of spatially separated graphene layers, where Dirac fermions are coupled only via many-body Coulomb interaction. Within the usual plasmon theory, based on the matrix Dyson equation for the Coulomb propagator in GMLS, we calculate the dispersion relations and Landau damping of multilayer plasmons and provide an understanding of their dependences on the inlayer electron density and the number of graphene layers, $N$. The plasmon modes obtained here for structures with a finite number of graphene layers are qualitatively different from the modes obtained previously in infinite graphene superlattices \cite{DasSarma2009} and in structures with a conventional parabolic spectrum of electrons\cite{DasSarma1982}. We find that the plasmon spectrum consists of one optical mode with a square-root dispersion and $N-1$ acoustical modes with linear dispersions in the bosonic momentum $q$ and in the long wavelength limit all these modes are undamped at zero temperature. In this limit we derive analytical formulas for the frequencies of multilayer optical and acoustical plasmon modes in GMLS. The frequency and the weight of the optical plasmons show an enhancement, respectively, as $\sqrt{N}$ and $N$, in full agreement with the measurements of Ref.~\onlinecite{Yan2012nnano}. We find that in GMLS the velocity of acoustical plasmons at vanishing momenta $q$ is given by a universal formula with different effective screening wave vectors, which differ from the Thomas-Fermi screening wave vector in an individual graphene sheet by numerical factors, determined by the total number of graphene layers $N$ and by the index of acoustical modes. Furthermore, we have presented detailed numerical calculations of the plasmon energy spectrum and the damping function in the full range of momenta in GMLS, consisting of up to $N=5$ graphene layers. Our comprehensive study includes finite temperature effects in different configurations of density balanced and unbalanced GMLS.

The paper is organized as follows. In Sec. II we present the model and use it to describe the multilayer plasmon modes in GMLS. Here we derive analytical expressions for the plasmon spectrum in $N$-layer graphene structures in the long wavelength limit. In Sec. III, we present our numerical calculations for arbitrary momentum and discuss the energy dispersions of the optical and acoustical plasmon modes and their corresponding damping as a function of $N$. In this section we study the effect of finite temperatures on the behavior of multilayer plasmon modes in different density balanced and unbalanced configurations of GMLS. In Sec. IV, we close the paper with a summary of our main results.

\section{Theoretical model and analytical results}

\begin{figure}[t]
\includegraphics[width=\columnwidth]{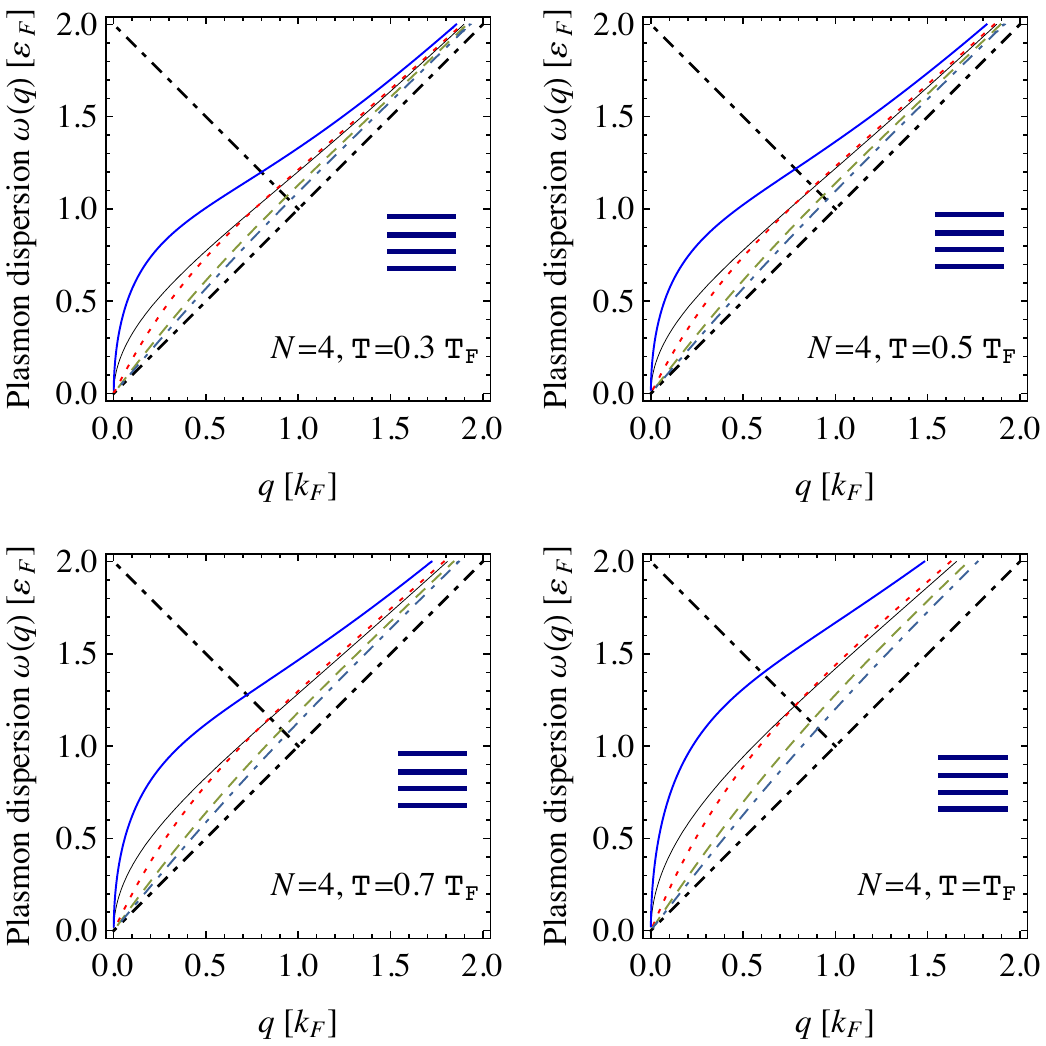}
\caption{(Color online) The plasmon dispersions in GMLS consisting of $N=4$ graphene monolayers for four different temperatures, $T=T_{F}$. All other parameters are the same as in Fig.~\ref{fig1}.}
\label{fig3}
\end{figure}

\begin{figure}[t]
\includegraphics[width=\columnwidth]{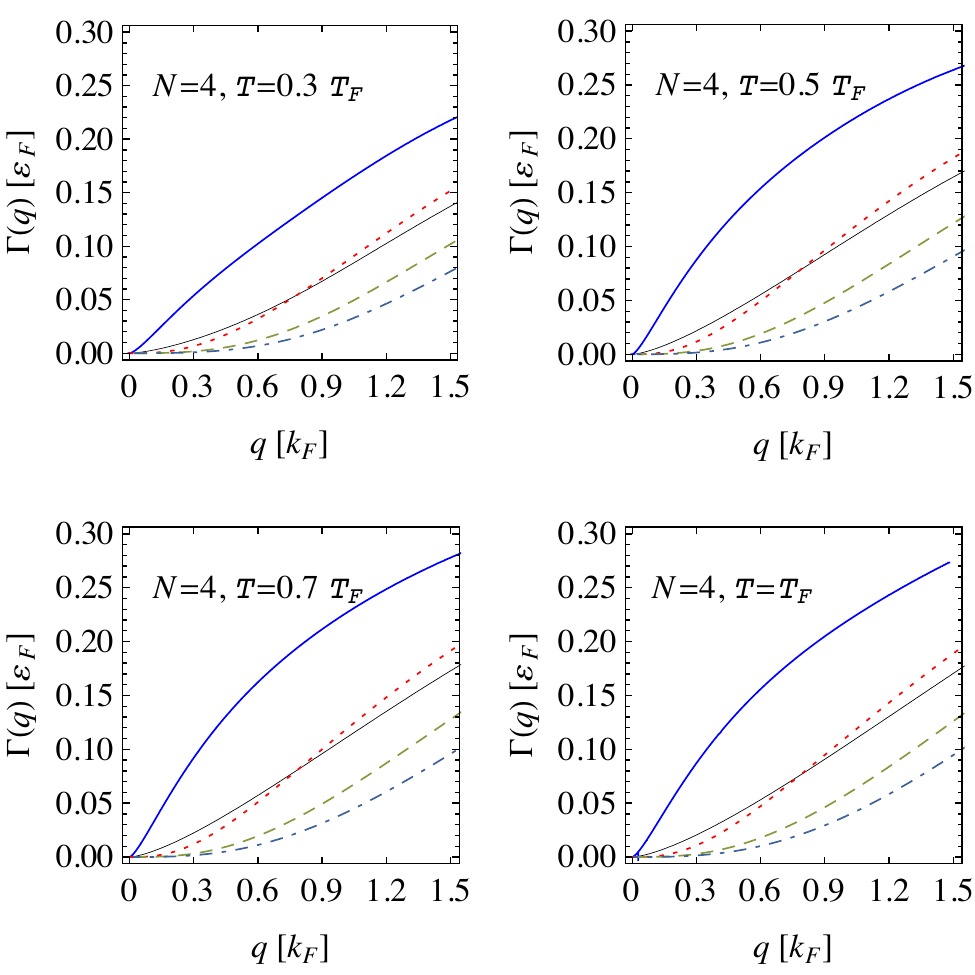}
\caption{(Color online) The broadening of the plasmon dispersions. The different curves here correspond to the multilayer plasmon branches with the same formatting and the same parameters as in Fig.~\ref{fig3}.}
\label{fig4}
\end{figure}

We consider GMLS consisting of $N$ Coulomb coupled graphene layers with an equal interlayer spacing $d$ and obtain the elementary excitations of the multilayer plasmon modes from the poles of the exact Coulomb Green function, $\hat{V}(q,\omega )$. In such GMLS the bare Coulomb interaction is a symmetric tensor 
\begin{equation}\label{BCI}
v_{ij}(q)=v(q) \exp \left(-|i-j| q d \right)
\end{equation}
with respect to the layer indices $i,j=1,2,\dots ,N$. Here $d$ is the interlayer spacing. The non-diagonal elements in Eq.~(\ref{BCI}) represent different interlayer electron-electron interactions while the diagonal elements are given by the Fourier transform of the intralayer Coulomb potential in two dimensional momentum space, $v(q)=2\pi e^{2}/\kappa q$. According to the experimental
situation of Ref.~\onlinecite{Yan2012nnano}, here we assume that the dielectric background of the environment surrounding graphene layers is homogeneous and can be well described by an average effective dielectric constant $\kappa $. The matrix kernel $\hat{v}(q)=v_{ij}(q)$ determines a standard matrix Dyson equation for the exact Coulomb Green function
\begin{equation}\label{Dyson}
\hat{V}(q,\omega )=\hat{v}(q)+\hat{v}(q)\cdot \hat{\Pi}(q,\omega )\cdot \hat{V}(q,\omega ) 
\end{equation}
where $\hat{\Pi}(q,\omega )$ is the irreducible polarization function of the GMLS. As seen from (\ref{Dyson}) the poles of the exact Coulomb Green functions are given by the zeros of the real part of the scalar screening function
\begin{equation}\label{DE}
\Re ~\varepsilon _{N}(q,\omega )=0
\end{equation}%
with
\begin{equation}\label{SFdet}
\varepsilon _{N}(q,\omega )=\text{det}\left|\mathbb{1}-\hat{v}(q)\cdot \hat{\Pi}(q,\omega )\right|~.
\end{equation}
In experiment the interlayer spacing in GMLS is sufficiently large so that the electrons in different spatially separated graphene monolayers are coupled only via many-body interlayer Coulomb interaction, \textit{i.e.} the electron tunneling between layers is insignificant and we can neglect the non-diagonal elements of the polarizability and take $\hat{\Pi}(q,\omega )=\delta _{ij}\Pi ^{i}(q,\omega )$. In the self-consistent random phase approximation \cite{GV} $\Pi ^{i}(q,\omega) $ is given in terms of the noninteracting Lindhard polarization function $\Pi_{0}^{i}(q,\omega )$, calculated quantum mechanically from the bubble diagrams in the $i$th graphene monolayer with an electron density $n_{i}$. This approximation neglects vertex corrections \cite{Abedinpour2011}, however, it usually provides a satisfactory description of the plasmon modes within the Fermi liquid formalism.

In balanced GMLS with an equal electron density in all monolayers, $n_{i}=n_{0}$, the screening function can be represented as a product of the form
\begin{equation}\label{DFsym}
\varepsilon _{N}(q,\omega )=\prod_{i=1}^{N}\left( 1-f_{N}^{i}(qd)v(q)\Pi_{0}(q,\omega )\right)
\end{equation}
where the set of $N$ different algebraic functions $f_{N}^{i}$ determine the actual dispersion relations of the plasmon modes in GMLS. An important feature of the Coulomb interaction is that for an arbitrary number of layers $N$ one of these functions $f_{N}^{1}$ in the long wavelength limit does not depend on the bosonic momentum $q$ and is given namely by the number of layers in GMLS,
\begin{equation}\label{f1}
f_{N}^{1}(qd)=N~.
\end{equation}
Accordingly, $f_{N}^{1}$ determines the frequency of the in-phase optical plasmon mode with a square root dispersion with $q$. Moreover, it can be shown that the remaining $N-1$ functions $f_{N}^{i}$ with $i=2,\dots,N$ vanish linearly with $q$ in this limit and can be represented as
\begin{equation}\label{fN}
f_{N}^{1+i}(qd)=\alpha^{i}_{N} q d~.
\end{equation}
These functions identify the $N-1$ out-of-phase acoustical plasmon modes with a linear dispersion. The numerical coefficients $\alpha^{i}_{N}$ with $i=1,\dots, N-1$ determine the velocity of the $i$th acoustical mode.

\begin{figure}[t]
\includegraphics[width=\columnwidth]{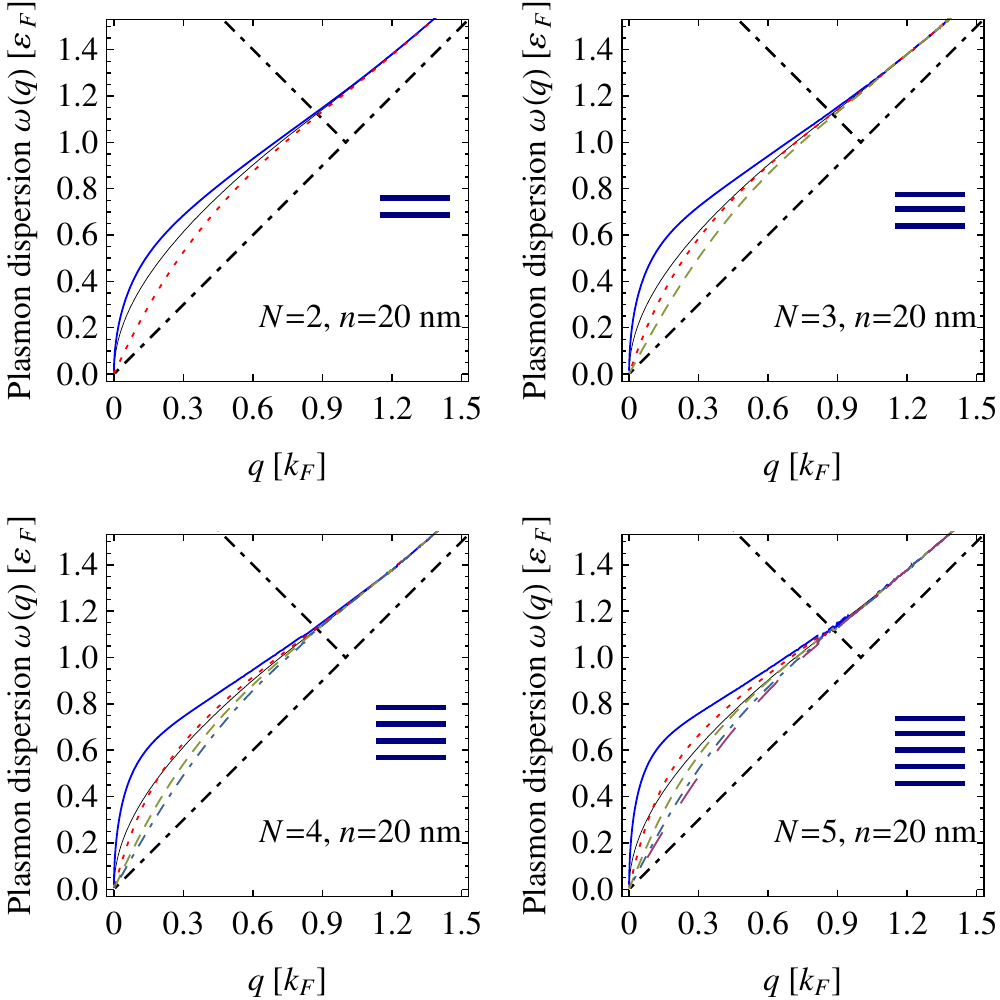}
\caption{(Color online) The plasmon dispersions at finite temperatures $T=0.1T_{F}$ in GMLS consisting of $N=2, 3, 4$ and $5$ graphene monolayers (shown in bold parallel lines in each panel) for the interlayer separation $d=20 $ nm. All other notations and parameters are the same as in Fig.~\ref{fig1}.}
\label{fig5}
\end{figure}


As seen from Eq.~(\ref{DFsym}) the effective polarizability for the in-phase optical mode is $\Pi _{in}=N\Pi_{0}(q,\omega )$. Its imaginary part determines the weight of the optical plasmon hence, in agreement with the experimental findings of Ref.~\onlinecite{Yan2012nnano}, it is additive in the number $N$ of monolayer graphene sheets in GMLS. This conclusion is independent of the quantum mechanical properties of the electron motion in each graphene layer that determines the properties of the Lindhard polarization function $\Pi _{0}(q,\omega )$ of single monolayer graphene. In the long wavelength limit the in-phase optical plasmon mode can be obtained from the high-frequency dynamical ($v_{F}q\ll {\omega }\ll {\varepsilon }_{F} $) asymptote of the Lindhard function, which can be written in the following universal form
\begin{equation}\label{PFdyn}
\Pi _{0}(q,\omega )\propto \frac{{\varepsilon }_{F}}{\hbar }\frac{q^{2}}{{\omega }^{2}}~.
\end{equation}%
In conventional two-dimensional electron gases with parabolic dispersion, the Fermi energy increases linearly with the doping level, ${\varepsilon }_{F}\propto n_{0}$, therefore the total in-phase polarizability given by Eq.~(\ref{PFdyn}) is additive in the {\it total} number of electrons, $\Pi _{in}\propto n_{tot}$ with $n_{tot}=Nn_{0}$, while the in-phase optical plasmon frequency ${\omega }_{op}\propto\sqrt{n_{tot}}$. The situation is strikingly different in GMLS. The unique spectrum of Dirac fermions with the constant velocity implies that the Fermi energy in a graphene monolayer, $\e_{F}=v_{F}k_{F}$ with $k_{F}=\sqrt{\pi n_{0}}$, is a square root function of the inlayer density $n_{0}$ hence the total in-phase polarizability of GMLS, again given by Eq.~(\ref{PFdyn}), is no longer additive in $n_{tot}$. However in virtue of Eq.~(\ref{f1}), it is still additive in the number of layers $N$ so that we have $\Pi _{in}\propto N\sqrt{n_{0}}$ and ${\omega }_{op}\propto \sqrt{N}{n_{0}}^{1/4}$, which is again in agreement with the measurements of Ref.~\onlinecite{Yan2012nnano}.

Because the dispersion of the acoustical plasmon modes is linear, one cannot \cite{SG1988,Profumo2012} use the dynamical ($\o/q\rightarrow \infty$) limit (\ref{PFdyn}) of the Lindhard function in the long wavelength limit in order to obtain the plasmon velocity. Instead, we obtain the dispersion of acoustical modes making use of the exact expression \cite{Hwang2007} for the zero temperature Lindhard polarization function in graphene. Thus, in the long wavelength limit we find the following square-root and linear energy dispersions
\begin{equation}\label{DR}
\omega _{op}(q)=\sqrt{Nq\frac{ge^{2}v_{F}k_{F}}{\kappa }}~,~\omega_{ac}^{i}(q)=\frac{1+\alpha^{i}_{N}q_{TF}d}{\sqrt{1+2\alpha^{i}_{N}q_{TF}d}}v_{F}q~,
\end{equation}
respectively, for one optical and $N-1$ acoustical multilayer plasmon modes. Here $g=4$ accounts for the spin and valley degeneracy in graphene and $q_{TF}=4 e^{2}k_{F}^{2}/(\kappa\varepsilon _{F})$ is the Thomas-Fermi screening wave vector in graphene. The numerical factors $\alpha^{i}_{N}$ ($i=1,\dots, N-1$) of the $i$th acoustical mode are uniquely determined by the total number of graphene layers $N$ in GMLS. For $N=2$ we have only one coefficient $\alpha^{1}_{2}=1$ and the above formulas recover the previous results obtained for the plasmon dispersions in double-layer graphene structures \cite{Hwang2009,Profumo2012,SMB2012}. For GMLS consisting of $N=3$ graphene layers, we find for the two coefficients 
\begin{equation}
\alpha^{1}_{3}=2~,~\alpha^{2}_{3}=2/3~,
\end{equation}%
corresponding to the two acoustical plasmon modes. For $N=5$ the four different velocities of the acoustical plasmon modes in Eq.~(\ref{DE}) are given in terms of the following coefficients
\begin{equation}
\alpha^{1,2}_{5}=\frac{4}{3\pm \sqrt{5}}~,~\alpha^{3,4}_{5}=\frac{4}{5\pm \sqrt{5}}~.
\end{equation}
In the limit of small $q_{TF}d$, the separations between $i$th and $j$th acoustical modes from the top of the intrasubband electron-hole continuum differ in the long wavelength limit by a factor $\gamma_{N}=(\omega^{i}_{ac}-v_{F}q)/(\omega^{j}_{ac}-v_{F}q)=\left(\alpha^{i}_{N}/\alpha^{j}_{N}\right)^{2}$. It gives a factor of $\gamma_{3} \approx 9$ difference for the acoustical  branches in GMLS with $N=3$ while for $N=5$ the difference for the most distant acoustical  modes is even larger, by a factor of $\gamma_{5}\approx90$. In experiment, however, the parameter $q_{TF}d $
is typically not very small and this difference is moderate. For the inlayer carrier density $n_{0}=10^{12}$cm$^{-2}$ and the interlayer spacing $d=5$nm taking $\kappa=3.8$, we have $q_{TF}d\approx 2$ and this gives $\gamma_{2}\approx 3$ while $\gamma_{5}\approx 7$.

\begin{figure}[t]
\includegraphics[width=\columnwidth]{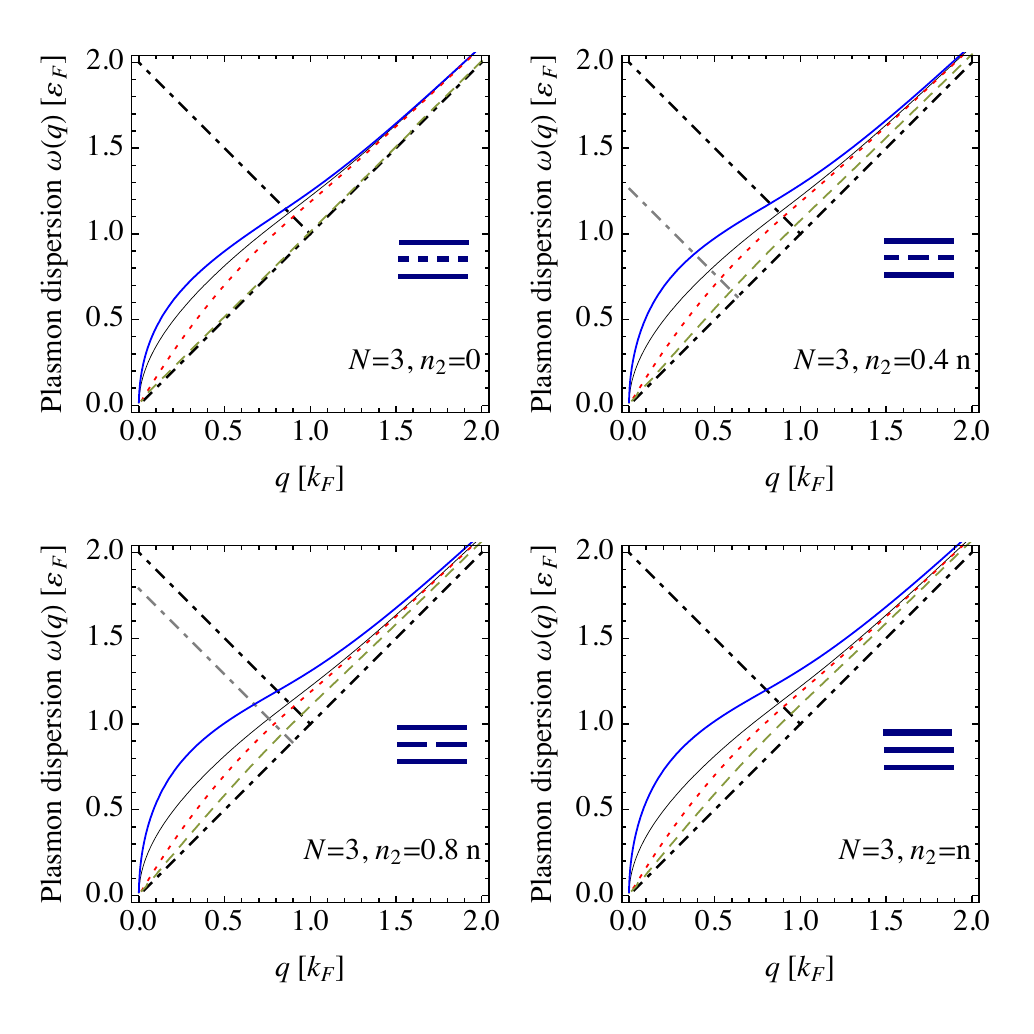}
\caption{(Color online) The plasmon dispersions in an unbalanced $N=3$-layer graphene structure. The different panels correspond to the inlayer carrier density $n_{i}=0, 0.4, 0.8, 1 \times 10^{12}$cm$^{-2}$ of the middle layer graphene sheet. All other parameters and notations are the same as in Fig.~\ref{fig1}. }
\label{fig7}
\end{figure}

\begin{figure}[t]
\includegraphics[width=\columnwidth]{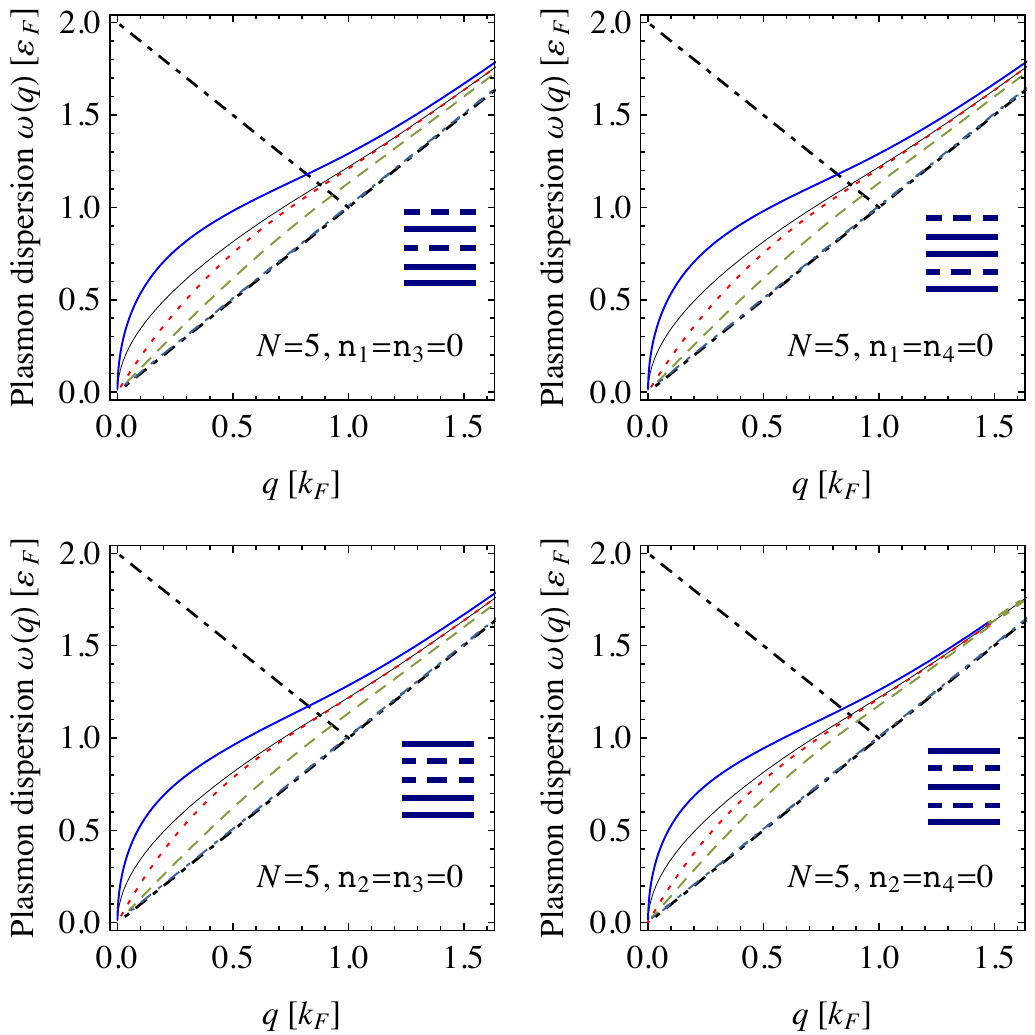}
\caption{(Color online) The plasmon dispersions in an unbalanced $N=5$-layer graphene structure with two unroped layers. The different panels correspond to the different situations of $n_{1}=n_{3}=0$, $n_{1}=n_{4}=0$, $n_{2}=n_{3}=0$, $n_{2}=n_{4}=0$. All other parameters and notations are the same as in Fig.~\protect\ref{fig1}. }
\label{fig8}
\end{figure}

\section{Plasmons dispersions in multilayer graphene structures}

In this section we present our numerical calculations for the energy dispersions and Landau damping of multilayer plasmon modes both in balanced and unbalanced GMLS with up to $N=5$ graphene layers. We calculate the energy spectrum from the general formula (\ref{DE}) of the dispersion equation with the screening function (\ref{SFdet}). For the diagonal elements of the polarization function $\Pi_{0}^{i}(q,\omega )$ we make use of the exact semianalytical formulas from Ref.~\onlinecite{Ramez} for the Lindhard polarization function and study also the effect of finite temperature both on the plasmon dispersion and damping. 
The Landau damping of the plasmon modes can be well described by the broadening function of the respective energy dispersions 
\begin{equation}
\Gamma _{i}(q)=\left. \Gamma (q,\omega )\right\vert _{\omega =\omega _{i}(q)}
\end{equation}%
where the function
\begin{equation}
\Gamma (q,\omega )=\frac{\Im \Pi ^{0}(q,\omega )}{\partial \Re \Pi
^{0}(q,\omega )/\partial \omega }
\end{equation}%
gives the property of the electron-hole continuum to cause damping of the elementary excitations. Thorough the calculations we assume that GMLS has a homogenous dielectric background with the dielectric constant $\kappa = 3.8$. In actual calculations of the dispersion relations we set also the imaginary part of $\Pi_{0}^{i}(q,\omega )$ to zero \cite{zimp} in the screening function (\ref{SFdet}), which provides a correct asymptote for the spectral branches of multilayer plasmons in the limit of $d \rightarrow \infty$.

In Fig.~\ref{fig1} we plot the dispersion relations of multilayer plasmons in graphene structures with $N=2, 3, 4$ and $5$ layers. The plasmon dispersion in a single layer is shown also by a thin line for reference. We see a clear shift of the in-phase optical plasmon mode towards higher energies with an increase of the number of graphene layers in GMLS. This enhancement is strongest at wave vectors around $q=0.3 k_{F}$ where we find an increase of the distance between the multilayer and the single-layer optical plasmon modes of about $0.17 \varepsilon_{F}$ for a double-layer structure, $N=2$, and of about $0.39\varepsilon_{F}$ for GMLS, consisting of  $N=5$ graphene layers. Strong modifications occur also for multilayer acoustical  plasmons. With an increase of $N$ we find an increase of the number of multilayer out-of-phase modes and also an enhancement in the energy of the upper lying acoustical modes. This enhancement is so strong that the top acoustical mode crosses the single layer plasmon mode in GMLS for $N=4$ and $5$ in the triangular region outside the electron-hole continuum. We find that in GMLS consisting of $N=5$ ($N=4$ and $N=3$) graphene layers, only $2$ ($1$) from the $4$ ($3$ and $2$) acoustical branches are suppressed as compared with the single acoustical mode in GMLS for $N=2$. The behavior of multilayer plasmon modes in the short wavelength limit remains almost unchanged. In this approximation, independent of the number of layers, all multilayer plasmon modes for $q\gg k_{F}$ coagulate around the single-layer plasmon branch and remain at some distance parallel to the boundary of the intra-chirality subband electron-hole continuum.

In Fig.~\ref{fig2} we study the Landau damping of the corresponding multilayer plasmon modes, shown in Fig.~\ref{fig1}. The energy enhancement, observed for all multilayer plasmon modes, accompanies with an enhancement of their damping in the number of layers $N$. It is seen that the damping is strong where the plasmon mode is far from the boundary of the intra-chirality subband electron-hole continuum hence the decay of the in-phase optical mode is always larger than that of the out-of-phase acoustical modes. This indicates that at finite temperatures the finite damping of the multilayer plasmon modes outside the electron-hole continuum is due to a leakage of the plasmon weight from the inter-subband electron-hole continuum towards low energies and momenta but from the intra-subband continuum, at which boundary, $\omega=v_{F}q$, the damping function vanishes, $\Gamma (q,v_{F}q )=0$. The broadening of a multilayer plasmon mode shows a strong increase at some upturn momentum, $q_{c}$, which is the smallest for the optical mode and the largest for the lowest in energy acoustical mode, respectively, with the values $q_{c}\approx 0.2 k_{F}$ and $0.8 k_{F}$ in GMLS consisting of $N=5$ layers. In the short wavelength limit, $q \gg k_{F}$, the broadening of multilayer plasmon modes, $\Gamma_{i}(q)$, increases approximately linearly in $q$, independent of the plasmon mode index $i$.

In Figs.~\ref{fig3} and \ref{fig4} we study the finite temperature effect on the plasmon dispersions and on the broadening of the respective plasmon modes in GMLS consisting of $N=4$ layers. It is seen that with an increase of $T$, the energy of all multilayer plasmon modes exhibits a significant increase. This enhancement results in a larger broadening of the corresponding plasmon dispersions because the distance of the respective modes from the boundary of the intrasubband electron-hole continuum increases. The other source of the broadening enhancement is the increasing leakage of the plasmon weight with $T$ from the intersubband  electron-hole continuum. The Landau damping increases drastically, especially for the in-phase optical and for the top out-of-phase acoustical plasmon modes so that the upturn momentum for these modes almost vanishes. It is seen that at temperatures $T\sim 0.5T_{F}$, the multilayer plasmon dispersions show up to the $15-20$ percent broadening in the triangular energy region outside the zero temperature electron-hole continuum.

The plasmon dispersions in GMLS consisting of up to $N=5$ graphene layers are shown in Fig.~\ref{fig5} for a larger value of the interlayer spacing, $d=20$ nm. One can see a clear convergence of the optical and all acoustical multilayer plasmon modes to the single-layer plasmon dispersion. This tendency is so strong that it is hard to distinguish between the multilayer plasmon modes when they enter into the inter-subband electron-hole continuum.


In Figs.~\ref{fig7}-\ref{fig8} we study the spectrum of the multilayer plasmon modes in unbalanced GMLS. Fig.~\ref{fig7} shows the plasmon dispersions in GMLS consisting of $N=3$ graphene layers for four different values of the inlayer density, $n_{2}$, in the middle layer graphene sheet. As seen in the top right panel for the vanishing density $n_{2}=0$ and finite temperatures $T=0.1T_{F}$, the lowest acoustic mode is slightly separated from the boundary of the intra-subband electron-hole continuum. With an increase of $n_{2}$ the separation increases along with an overall enhancement of the energy of all multilayer plasmon modes. In Fig.~\ref{fig8} we calculate the plasmon dispersions in GMLS consisting of $N=5$ graphene layers. The dopping level in two layers of them is at the Dirac point. It is seen that accordingly only two from the four acoustical modes are well separated from the boundary of the intra-subband electron-hole continuum at low temperatures $T=0.1T_{F}$. Notice that the energy of all multilayer plasmon modes depends weakly on the position of the pristine graphene layers. In the configuration shown in the bottom left panel, the optical plasmon is slightly suppressed while the acoustical modes are slightly enhanced in energy as compared with the plasmons in the other configurations.   

\section{Summary}

We have theoretically investigated the collective plasma oscillations in $N$-layer graphene structures where Dirac Fermions in different spatially separated layers are coupled via Coulomb interaction. We have calculated the energy dispersions and Landau damping of the multilayer plasmon excitations by changing the total number of layers, the inlayer carrier density, the interlayer spacing, the lattice temperature, and the number and positions of undopped layers in such graphene structures. The multilayer plasmon spectrum consists of one optical plasmon mode with a square-root dispersion and $N-1$ acoustical plasmon modes with linear dispersion.  We have found that in the long wave length limit and for any number of graphene layers in GMLS, the energy of the optical plasmon mode and its weight exhibit, respectively, a square-root and a linear enhancement with $N$. This calculated behavior is in agreement with the recent experimental findings of Ref.~\onlinecite{Yan2012nnano} and provide a detailed understanding of the observed enhancement effect. The velocity of the upper lying acoustical branches of the multilayer plasmon spectrum increase with the number of graphene layers. The obtained enhancement is largest for the top outermost acoustical mode, which crosses at some value of the wave vector, $q\sim k_{F}$, the plasmon energy of an individual graphene sheet.  The velocity of the low lying acoustical branches exhibits a weak suppression in graphene structures consisting of $N\geq 3$ layers as compared with the single acoustical mode in double-layer graphene structures. The presented numerical calculations provide an understanding of the finite temperature effect on the plasmon dispersions, their behavior versus the interlayer separation and the inlayer doping level. 


\acknowledgements

This work was supported by the ESF-Eurocores program EuroGRAPHENE (CONGRAN project) and the Flemish Science Foundation (FWO-Vl). S.M.B. acknowledges support from the Belgian Science Policy (BELSPO) and EU.

\end{document}